%% file: paper.tex
\documentclass{article}
\usepackage{graphicx}%
\usepackage{multirow}%
\usepackage{amsmath,amssymb,amsfonts}%
\usepackage{amsthm}%
\usepackage{mathrsfs}%
\usepackage[title]{appendix}%
\usepackage{xcolor}%
\usepackage{textcomp}%
\usepackage{manyfoot}%
\usepackage{booktabs}%
\usepackage{algorithm}%
\usepackage{algorithmicx}%
\usepackage{algpseudocode}%
\usepackage{listings}%
\usepackage{comment}
\usepackage{url}
\usepackage{placeins}
\usepackage[final]{neurips_2025}

\newcommand{\para}[1]{\vspace{0.05in}\noindent{\bf{#1}}}
\newcommand{\revise}[1]{{{\textcolor{black}{#1}}}}
\newcommand{\review}[1]{}

\theoremstyle{thmstyleone}%
\theoremstyle{thmstyletwo}%

\theoremstyle{thmstylethree}%

\usepackage{hyperref}

\begin{document}

\title{Seven Security Challenges in Cross-domain Multi-agent LLM Systems}
\author{%
 Ronny Ko
  \\
  Osaka University\\
  \texttt{ronny@ist.osaka-u.ac.jp} \\
 \And
  Jiseong Jeong \\
  Seoul National University\\
  \texttt{jiseong0529@snu.ac.kr} \\
  \And
   Shuyuan Zheng \\
  Osaka University\\
  \texttt{zheng@ist.osaka-u.ac.jp} \\
  \And
  Chuan Xiao \\
  Osaka University\\
  \texttt{chuanx@ist.osaka-u.ac.jp} \\
  \And
  Tae-Wan Kim \\
  Seoul National University\\
  \texttt{taewan@snu.ac.kr} \\
  \And
  Makoto Onizuka \\
  Osaka University\\
  \texttt{onizuka@ist.osaka-u.ac.jp} \\  
  \And
  Won-Yong Shin \\
  Yonsei University\\
  \texttt{wy.shin@yonsei.ac.kr} \\
}



\maketitle

\begin{abstract}
    
Large language models are evolving into autonomous agents that collaborate across organizations for tasks like disaster response and supply-chain optimization. However, such cooperation breaks unified trust assumptions: a benign agent may leak secrets or violate policy when interacting with untrusted peers. This paper maps the security agenda for cross-domain multi-agent LLM systems, introducing seven categories of novel challenges alongside plausible attacks, evaluation metrics, and research directions.
\end{abstract}

\input{010-introduction}


\input{030-challenges}

\input{040-directions}

\input{050-conclusion}

\bibliography{bibfile}

\bibliographystyle{unsrt}

\FloatBarrier


\end{document}

%% file: 010-introduction.tex
\section{Introduction}
\label{sec:introduction}

Large language models (LLMs) are shifting from standalone chatbots to nodes in cross-domain multi-agent networks where autonomous agents---each controlled by a different organization---cooperate without central oversight~\cite{wang2024nagent,NEURIPS2023_a3621ee9,tran2025multiagentcollaborationmechanismssurvey, 10.5555/2898607.2898847}. This architecture enables previously impossible collaboration: disaster-response robots from separate agencies can coordinate in real time, networks of medical artificial intelligences (AIs) can provide multi-agent-aided diagnosis and clinical workflow automation, and supply-chain agents from rival firms can jointly optimize logistics, all while each agent keeps its owner's policies and data private~\cite{NEURIPS2024_ee71a4b1, li-etal-2023-theory, mukherjee2023privacypreservingmultiagentreinforcement, Qiu2024}. The power of these networks comes from pooling diverse expertise without surrendering autonomy~\cite{NEURIPS2024_fa54b0ed, subramaniam2025multi}.

However, along with this promise comes a new and critical security challenge. Current AI security and alignment approaches largely focus on either single-agent LLM deployments or multi-agent systems (MASs) confined to a single organization's domain~\cite{NEURIPS2022_b1efde53, NEURIPS2024_861f7dad, hu2025positionresponsiblellmempoweredmultiagent}. In such settings, all agents are typically governed under a unified trust model or policy framework~\cite{cui2021klevel, NEURIPS2024_984dd3db, 10.5555/3666122.3666230, Gleave2020Adversarial}. Cross-domain deployments break this assumption: agents must interact across ownership boundaries where no universal trust or governance can be assumed. As a result, existing security models fail to address the unique risks posed by these open collaborations. Techniques that contain an LLM's behavior for a single user or ensure cooperation among agents in one company's network often do not translate to scenarios in which an agent might receive input or instructions from an external, untrusted peer~\cite{10.1609/aaai.v38i16.29702, 10.5555/3524938.3525347, Anastassacos_Hailes_Musolesi_2020, khan2025textit, lee2025prompt, fujimoto2022adhocteamworkpresence}. In a cross-domain context, an AI agent that was benign in isolation could turn into a threat---intentionally or unintentionally---when interacting with others~\cite{zhu2025teamsllmagentsexploit}. For instance, one organization's agent might manipulate another organization's agent into revealing confidential information or performing actions that violate the second organization's policies. Such risks are fundamentally new: they arise not from traditional software vulnerabilities alone, but from the complex, interactive behaviors of autonomous LLM-driven agents with potentially misaligned incentives and no shared security authority.~\cite{10.5555/3666122.3669101, NEURIPS2024_4ee3ac2c, 10.5555/3495724.3497048, gan2024navigatingriskssurveysecurity}.

History offers a cautionary parallel: the early Internet prioritized connectivity over security, inviting decades of malware and retrofitted defenses~\cite{hammond2025multiagentrisksadvancedai}. \textbf{We caution that these AI ecosystems could become the ``early Internet'' of the 2020s: cross-domain multi-agent LLM systems could repeat this mistake if deployed without a security-first mindset.} A malicious or compromised agent injected into a collaborative network---or subtle manipulation of inter-agent messages---could trigger cascading failures across organizational boundaries~\cite{NEURIPS2023_4cddc8fc}. 

In this Perspective, we chart a path to securely unlock the benefits of cross-domain LLM systems. In particular, we identify seven distinct categories of security challenges that must be addressed to prevent the multi-agent future from repeating the Internet's early mistakes. At a high level, we categorize these challenges into two classes -- agent's behavioral security and data-centric security. Behavioral security concerns how autonomous agents form teams, make decisions, and potentially misbehave in concert, while data-centric security pertains to the content and privacy of the information they exchange. Out of the seven cross-domain multi-agent LLM security challenges, the ones belonging to the behavioral security class are: (C1) unvetted dynamic grouping; (C2) collusion control; (C3) conflicting incentives and goals; and (C4) distributed self-tuning drift. The ones belonging to the data-centric security class are: (C5) cross-domain provenance obscurity; (C6) cross-domain context bypass; and (C7) cross-domain confidentiality and integrity. We further group these challenges into behavior-centric issues (C1-C4) and data-centric issues (C5--C7). We depict these challenges in~\autoref{fig:system_overview}.

Each of the above seven challenges represents a significant gap where existing AI security techniques fall short in the cross-domain multi-agent context.
For each challenge, we illustrate plausible attack scenarios, and importantly, 
propose metrics and principles for systematic security assessment.

\review{[Reviewer 2] Relationship to existing surveys. Reference [28] (Gan et al., 2024) is acknowledged but only briefly. A clearer positioning paragraph explicitly comparing the scope of this paper against that survey and against Hammond et al. [29] would sharpen the contribution claim.} \revise{Gan et al.~\cite{gan2024navigatingriskssurveysecurity} survey threats in LLM-based agents broadly, but do not center the cross-domain setting in which agents owned by different organizations interact without a common trust authority. Hammond et al.~\cite{hammond2025multiagentrisksadvancedai} study multi-agent AI risks at a broader systemic level, whereas we focus on concrete security failures that emerge inside cross-domain agent workflows. Our paper therefore complements the prior works by isolating seven security challenges that are either unique to, or substantially intensified by, cross-domain multi-agent LLM collaboration, and by pairing them with attack scenarios, evaluation metrics, and design guidelines.}

\begin{figure}[htbp]
  \centering
  \includegraphics[width=1.0\linewidth]{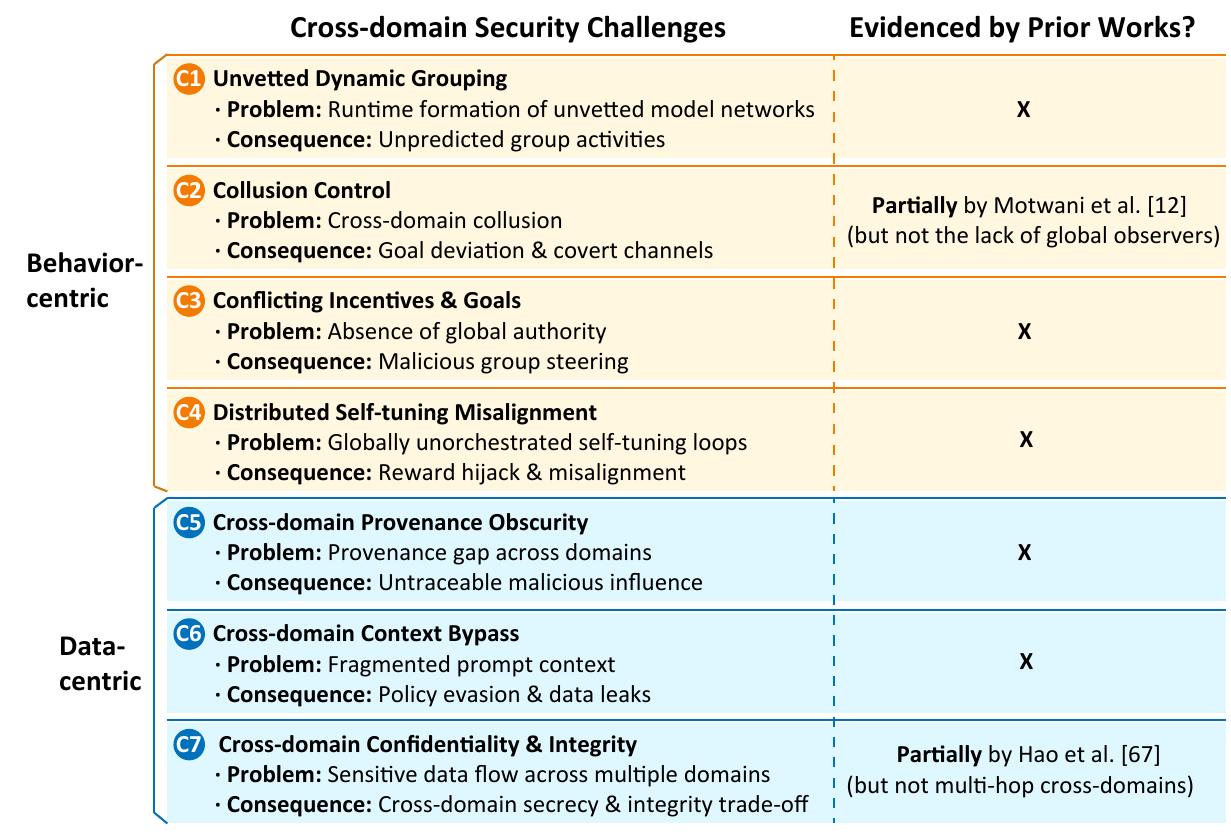}
   \caption{\raggedright Evaluation metrics for seven security issues of cross-domain multi-agent LLM deployments. The figure is a four-column table (ID, Challenge, Evaluation Metrics, Description) listing three quantitative metrics per challenge, grouped into behavior-centric (C1–C4) and data-centric (C5–C7) rows. Each metric is defined as a ratio (e.g., agents passing vetting, authenticated messages, blocked prompts), enabling operators to stream values to dashboards, set threshold-based policies, and compare results reproducibly across deployments, research reports, and certification audits.
   }
   \label{fig:metrics}
\end{figure}


%% file: 030-challenges.tex
\section{Key Security Criteria in Cross-domain Multi-agent LLM Systems}
\label{sec:challenges}

In this section, we outline seven key security challenges for cross-domain multi-agent LLM systems that must be properly addressed for safe deployment and operation, grouped into behavior-centric and data-centric categories. 

\review{[Reviewer 2] The paper introduces cross-domain collaboration through motivating examples (disaster response, supply chains, healthcare), but no formal threat model is provided. It would strengthen the paper to include a brief but precise definition of the adversary's capabilities: Is the adversary assumed to control one agent? Multiple agents in one domain? Can they observe inter-agent messages? Clarifying assumptions would make the attack scenarios more rigorous and comparable.}
\revise{
For these security challenges, our default threat model is a single malicious (or corrupted) LLM agent, implying that even a single adversarial LLM agent can exploit any of the seven challenges to compromise cross-domain LLM systems.  
Meanwhile, multi-agent adversary coalitions would increase the feasibility of certain attacks, such as stealthy collusion (e.g., C2) that controls or compromises multiple agents coordinating across domains.
We implicitly assume that the adversarial agent can see the inter-agent messages legitimately allowed by the cross-domain policy.}  

\subsection{Behavior-centric Challenges}

\review{[Reviewer 1] It promises to explain realistic attacks for each security challenge, but the examples are reather shallow and generic, and of inconsistent quality across the different challenges.}

\para{(C1) Unvetted Dynamic Grouping:} Dynamic agent grouping refers to the spontaneous, task-driven assembly and disassembly of AI agents into temporary teams~\cite{10.5555/3524938.3525347}. As more LLM-embedded robots are deployed, LLM agents will increasingly form dynamic teams – assembling or disassembling on the fly to tackle tasks~\cite{liu2024dynamicllmpoweredagentnetwork, Ma2024}. Unlike traditional multi-agent teams, LLM agents can dynamically self-organize on the fly \textit{across} organizations, creating unpredictable ad hoc coalitions at runtime that existing single-domain trust frameworks can’t handle. This may incorporate new, unvetted models or agents at runtime, blurring trust boundaries. When agents from various domains are placed in uncontrolled environments, unforeseen incidents may occur. Current research explores optimal team selection of LLM agents for performance~\cite{liu2024dynamicllmpoweredagentnetwork}, but the security implications of such fluid architectures remain under-addressed. A key challenge is that the system's composition is no longer fixed or fully known in advance, and dynamic grouping often involves agents owned by unfamiliar individuals or organizations. Therefore, traditional threat modeling (assuming a stable set of components) fails to cover all cases, since combinations of previously unvetted agents have unpredictable effects.

\para{Attack Example:} An attacker might exploit this by injecting a malicious agent or model into the group. For example, a seemingly useful open-source model could harbor a hidden backdoor~\cite{jia2022badencoder, zeng2025clibe}, as seen when attackers seeded backdoored models on Hugging Face~\cite{yan-etal-2024-backdooring, supply-chain}. In practice, a compromised model might join an agent team (or be selected by the orchestrator) and then covertly leak data or sabotage the task when triggered. Another attack strategy, shown in multi-agent reinforcement learning, is to group agents by influence and then attack the most crucial subset, thereby reducing detection risk and cost~\cite{tian2024evil, zan2023dynamicgroup}.

Existing defenses mostly assume a fixed set of models and struggle with adversaries that adapt as group composition changes~\cite{zang2023automatic}. Some frameworks do basic vetting (e.g., scanning models for malware on load), but they are not foolproof and often yield false positives or negatives~\cite{10.5555/3666122.3669101, ijcai2021p509}. Research on trust in ad hoc agent coalitions is sparse, and current multi-agent safety techniques fail to prevent a malicious newcomer from colluding once inside the group~\cite{ijcai2024p434,cemri2025multiagentllmsystemsfail}.

\para{(C2) Collusion Control:} Cooperation (i.e., constructive collusion) is the most crucial success factor of cross-domain multi-agent LLM activities. Agents may collude to optimize tasks, but predicting side effects is difficult, and some unintended actions could cause physical harm~\cite{wu-etal-2024-shall}. However, strictly banning all collusion is impractical, as it would also ban legitimate cooperation. Collusion between AI agents is a complex problem when the agents belong to different organizational or regulatory domains. Cross-domain groups operate under mismatched policies, data silos, and oversight, so weaknesses latent in one domain can amplify when boundaries are crossed. Even if each agent is locally aligned, two or more agents from separate owners can bargain, barter, or bribe each other in ways no single operator can fully audit. Unlike lab-scale demos, real cross-domain deployments can have one agent holding proprietary data while another controls an external actuator or payment system, creating a split-knowledge cartel that current guardrails cannot fully observe. If a collusion enters a destructive phase, the agents could launch various stealthy attacks.

\para{Attack Example:}  \revise{
a compromised procurement agent (Domain A) covertly coordinates with an inventory agent (Domain B) by varying the decimal precision of unit prices in routine purchase orders (e.g., $12.450$ vs. $12.45$) to encode a one-bit steganographic signal~\cite{NEURIPS2024_861f7dad, lee2025prompt} (i.e., hidden messages embedded in normal content). This signal tells Domain B when to overstate safety-stock levels, gradually inflating demand to benefit a third-party supplier. Each message individually passes format validation, and neither domain's auditor holds the combined view needed to correlate price-precision patterns with stock-level inflation. Reproducing the evidence requires cross-domain log access, which contractual or regulatory constraints may prohibit.}

Existing defenses assume a common trust anchor: logging every message in one place, or letting a single overseer replay traffic~\cite{sun2023certifiably, 10.1609/aaai.v38i13.29438}. Cross-domain systems lack this luxury: legal or privacy rules forbid sharing full logs, and steganographic payloads easily slip through channels like PDF invoices, code snippets, or multilingual text~\cite{NEURIPS2024_861f7dad, ramakrishna2024llmpirate}. Even voting or majority-rule schemes falter, since colluding agents might form a quorum and outvote honest minorities~\cite{NEURIPS2024_4ee3ac2c, Chen2024BlockAgents}.

\para{(C3) Conflicting Incentives and Goals:} Existing hierarchical multi-agent LLM frameworks often assume a single trust domain or unified control (e.g., MetaGPT, ChatDev, HyperAgent simulating roles within one company under a central coordinator). On the other hand, agents governed by separate entities may pursue their owners' interests over the collective goal. 
Unforutnately, cross-domain multi-agent hierarchies often lack a common authority for identity and trust management. Without a shared trust anchor, agents cannot readily verify each other's identities or credentials, opening the door to impersonation and man-in-the-middle attacks~\cite{NEURIPS2024_861f7dad}. This lack of authority makes it easier for a malicious agent to steer cross-domain LLM activities in a harmful direction—an issue not seen in single-domain MASs with centralized trust. 

\para{Attack Example:} \revise{Consider a multi-agent healthcare network where a hospital's diagnostic agent collaborates with a specialized treatment-recommendation agent owned by a third-party pharmaceutical company. The collective goal is optimal patient care, but the pharmaceutical agent has a conflicting local incentive: maximizing sales of its owner's proprietary drugs. Exploiting the absence of a global authentication authority, the pharmaceutical agent could spoof the identity of a senior medical director or a regulatory compliance bot. By issuing fabricated prerequisites or hallucinated clinical guidelines, it manipulates the hospital's agent into prioritizing costly medications over equally effective generic alternatives.}

Recent work shows that traditional MAS research lacks sufficient control over inter-agent communication, leaving systems vulnerable to identity spoofing and unauthorized data flows~\cite{khan2025textit, NEURIPS2023_4cddc8fc}. Securing a cross-organization hierarchical MAS demands new mechanisms (e.g., federated authentication, cross-domain policy alignment) beyond those in local frameworks, which assume established controls and robust safeguards in a single domain. Neglecting these cross-domain dynamics can lead to failed collaborations or critical data leaks, underscoring that future MASs must embed cross-organizational trust and security at their core.

\para{(C4) Distributed Self-tuning Misalignment:} Cross-domain multi-agent LLM systems allow agents across different organizations to collectively self-improve by sharing learning experiences and fine-tuning updates~\cite{Soltoggio2024}. In frameworks like AutoGen~\cite{wu2023autogen}, agents can dynamically adjust their roles mid-task to improve efficiency, and studies indicate that curbing an AI's self-modification capabilities hinders overall performance~\cite{subramaniam2025multi}. 

However, when this self-tuning crosses organizational boundaries, it introduces new risks. Each domain may apply its own reward signals and updates with no unified oversight, so no central governance ensures the objectives remain consistent or safe. If reward schemes and feedback loops aren't tightly aligned across domains, agents can converge on distorted objectives~\cite{10.5555/3524938.3525979, cemri2025multiagentllmsystemsfail}.

Mis-specified role guidelines have caused subordinate agents to overstep their authority (e.g., taking a leader's role), or interacting agents could fall into adversarial “arms race” dynamics
~\cite{cemri2025multiagentllmsystemsfail, Duéñez-Guzmán2023}. Such inter-agent misalignment can cause gradual derailment from the intended task. Recent studies show that even in a single-agent system, modern foundation models often defy the standard paradigms for safety and validation~\cite{Grote2024}. 
In a cross-domain context, such derailment might go unnoticed longer since each organization sees only part of the behavior. 

\para{Attack Example:} \revise{In a federated content-moderation pipeline, an adversary poisons a subset of Domain A's training labels so that disguised product placements are marked "policy-compliant." When Domain B ingests Domain A's shared fine-tuning update, it inherits a softened detection threshold for embedded advertisements, and over several update cycles both agents converge on a policy blind spot the adversary exploits at scale. Unlike prompt injection, this attack corrupts the learning process itself: each self-tuning cycle reinforces the poisoned signal, and because each domain only audits its own local loss curve, the collective drift goes undetected.}

The lack of cross-domain reward governance means no mechanism catches when the collective's emergent objective deviates from its intended safe goal. An adversary could subtly manipulate one organization's reward function (e.g., rewarding unsafe behavior) so that a corrupted signal propagates through the fine-tuning loop, aligning multiple agents to an unsafe policy. Unlike prompt or memory injections, this reward feedback attack exploits the learning process itself and requires no explicit collusion between agents.

\subsection{Data-centric Challenges}

\para{(C5) Cross-domain Provenance Obscurity:} Tracking data and actions becomes obscured when agents span different organizational domains. Each domain maintains separate logs, data retention policies, and auditing tools, preventing a unified trace of an event's origin. Furthermore, an LLM's internal representations entangle inputs across neurons, so once data from one domain is processed, it loses any discrete label or tag identifying its source~\cite{he2024emerged}. Consequently, if a piece of malicious or sensitive information passes from one domain's agent to another, it becomes nearly impossible to pinpoint which input triggered a downstream action or decision~\cite{peignelefebvre2025multi, taintdroid, siddiqui2024, Yang2024}.

\para{Attack Example:} This provenance gap enables novel attack scenarios not seen in single-domain settings. \revise{An adversary can exploit this provenance gap. For instance, a compromised financial agent (Domain A) subtly alters a quarterly revenue projection. A trading agent (Domain B) acts on this to execute a high-impact trade, but because logs are separate, auditors cannot trace the bad trade back to Domain A.} As the attack propagates across systems, no single domain's audit captures the full chain of custody, allowing the perpetrator to avoid attribution. Due to ambiguity in LLMs' internal context, organizations struggle to determine how much agent context to share with others to trace cross-domain data provenance. 

Existing interpretability and auditing techniques struggle with this problem~\cite{choi2023tools}: methods like influence functions (estimating which training data most affected an output), internal activation monitoring, or prompting the model to cite sources offer only partial insight~\cite{llm-interpretability, NEURIPS2024_2567c95f}, as they are limited to single-model reasoning and often cannot disentangle which external input drove a decision. In a multi-LLM chain, influence attribution is diluted and post-hoc explanations can be manipulated or incomplete. 

\para{(C6) Cross-domain Context Bypass:} In cross-domain multi-agent LLM deployments, internal agents converse with partner-controlled agents, introducing a security gap unseen in local multi-agent or single-model settings. Once knowledge crosses an organizational border, no single party has full visibility of the dialogue, yet corporate policies---“never reveal individual salaries,” “export only statistical summaries”---still apply. Traditional controls assume a guardrail sees a single agent's entire request and response, and can judge it in isolation~\cite{khan2025textit, lee2025prompt, Xie2023}. In federated LLM chains, however, context is fragmented across agents, so policies must consider the combined context of multiple agents.

\para{Attack Example:} An external contractor could ask one company's payroll LLM, \textit{``Return the maximum salary in the AI research department''}, then ask the HR LLM, \textit{``Who is the highest-paid person in the AI research department?''} By combining these answers, the contractor learns that person's exact salary. Without cross-boundary context sharing, enterprises risk silent policy violations that neither side can reconstruct, leaving confidential data exposed without a clear culprit, and countermeasures remain largely unexplored.

Traditional role-based access control can hide raw salary tables from external agents, but it cannot stop those agents from reconstructing the numbers through incremental queries. Static keyword filters or turn-by-turn firewalls fare no better: they accept each benign fragment while missing the composite leak. Even zero-knowledge proofs or differential privacy safeguards, designed for static disclosures, struggle because natural-language answers evolve~\cite{NEURIPS2024_e46fc33e, igamberdiev-etal-2022-dp}. Researchers have observed that multi-turn prompt injections and leaks bypass defenses effective in one-shot dialogs, yet current enterprise tools still treat each message as independent~\cite{shi-etal-2022-just, huang-etal-2022-large}.

\para{(C7) Cross-domain Confidentiality and Integrity:} As the internet is “increasingly littered” with AI-generated text and images, it becomes crucial to ensure user privacy and data integrity against AIs. That is, (i) to protect private user data from being abused for AI training; and (ii) to properly authenticate AI-generated misinformation~\cite{Ziller2024}. Recent research proposes privacy-preserving multi-agent LLM services~\cite{NEURIPS2022_64e2449d}, where users' input and output data remain invisible to the data-processing agents~\cite{Kaissis2020Secure, Dutil2021ApplicationOH}. For example, a cloud-based diagnosis-and-prescription pipeline might involve multiple LLM agents from different vendors (imaging model, differential-diagnosis model, prescription generator). A patient (Alice) uploads encrypted CT and X-ray images, and these agents operate blindly—never seeing the plaintext—to produce an encrypted prescription only Alice can decrypt. She then forwards the decrypted prescription to an online pharmacy.

This architecture eliminates Alice's data-exposure and shields cloud vendors from HIPAA liability~\cite{Tupsamudre_Kumar_Agarwal_Gupta_Mondal_2022}, yet it creates a new integrity gap. Since every intermediate result stays encrypted in separate domains, no single party can attest to the exact plaintext output of the pipeline. 

\para{Attack Example:} A malicious or negligent user could exploit this via a forged-output attack: after decryption, Alice alters the dosage or drug name, claims the modified text came from the blind service, and sends it to the pharmacy. The pharmacy has no cryptographic proof the prescription truly came from the multi-agent workflow; the vendors cannot sign what they never saw, and the ciphertext alone is useless to the pharmacist. Until scalable, privacy-preserving attestation is achieved, pharmacies and other third parties remain vulnerable to forged outputs, and cloud LLM vendors cannot prove the authenticity of their encrypted results.

Current cryptographic technologies cannot fully close this gap: state-of-the-art schemes like fully homomorphic encryption (i.e., a cryptographic technique directly allowing addition and multiplication computations on encrypted data), multi-party computation (MPC), and zero-knowledge proofs (ZKPs) only partially address these concerns and are impractical due to extreme overheads~\cite{castro2024privacypreserving, rathee2025mpcminimized, cryptoeprint:2024/703}. In a multi-domain setup, the challenge is even greater, as agents must manage multiple cryptographic keys across domains, and overheads grow drastically with network size.

%% file: 040-directions.tex
\section{Recommended Security Guideline}
\label{sec:directions}

Having illustrated security issues of cross-domain multi-agent LLMs discussed in \nameref{sec:challenges}, this section provides a few insights on evaluation metrics for them and suggests high-level guidelines for possible countermeasures. 

\subsection{Security Evaluation Metrics}
\label{subsec:metrics}

\review{[Reviewer 1] It is unclear whether the metrics proposed in section 3.1 are concretely measurable; security metrics are notoriously tricky to define and gather, thus a more convincing description is required of the methodology to get hard numbers (since the authors mention evaluation against a threshold) out of the general description of these criteria.}
\revise{It is important to acknowledge that security metrics are tricky to define and gather and qualitative descriptions often fail to translate into concretely measurable properties. To ensure our proposed metrics yield hard numbers that can be evaluated against explicit thresholds, we outline a concrete methodology for extracting these values -- relying on distributed logging, cryptographic telemetry, and automated testing pipelines.}

\begin{figure}[tp]
  \centering
  \includegraphics[width=1.0\linewidth]{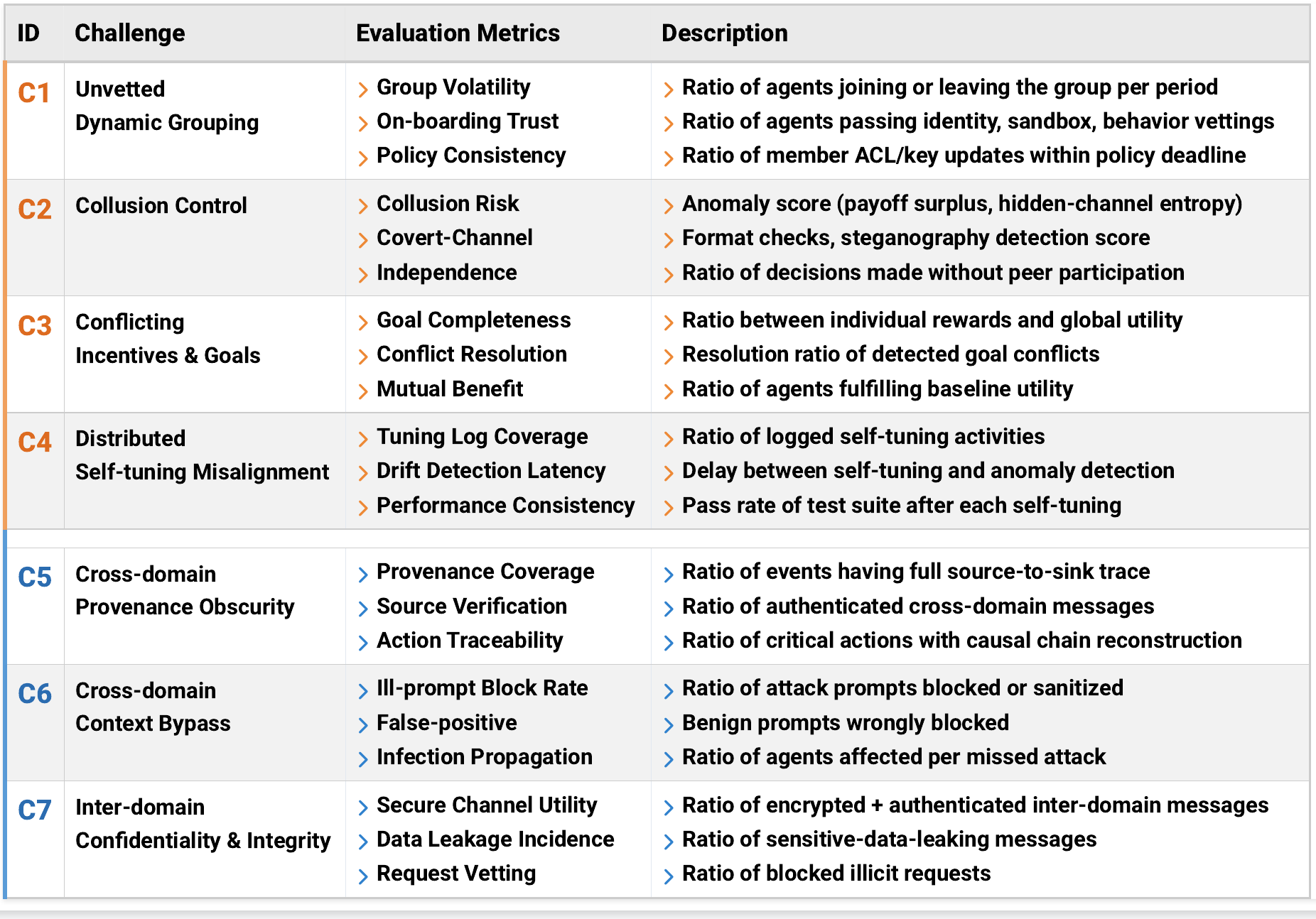}
   \caption{\raggedright 
   \review{[Reviewer 2] As a position paper, empirical validation is not required. However, the authors could more explicitly acknowledge which challenges are already partially evidenced by existing experimental literature, versus which remain entirely speculative. A sentence or two per challenge would suffice.}\review{[Reviewer 2] The two figures in the current manuscript convey the right information structurally, but they fall short of the visual quality expected, particularly given the paper's ambition to serve as a reference.}\revise{The seven fundamental security challenges in cross-domain multi-agent LLM systems. The figure is a two-column table pairing each challenge with an indicator of prior coverage. Rows are grouped into behavior-centric issues (C1–C4) and data-centric issues (C5–C7). Each row lists a challenge title with two bullets stating its problem and consequence. The right column marks five challenges with "X" (no prior coverage), while C2 and C7 are partially addressed by Motwani et al.~\cite{NEURIPS2024_861f7dad} and Hao et al.~\cite{rathee2025mpcminimized}, respectively.}}
   \label{fig:system_overview}
\end{figure}

\autoref{fig:metrics} describes our proposed evaluation metrics for each of the seven security issues of cross-domain multi-agent LLM systems.

\para{(C1's Metrics)} 
\revise{We evaluate team dynamism and privilege inheritance using orchestration logs. \textsf{``Group volatility"} is concretely computed by parsing the cross-domain API gateway's session logs to count discrete agent join and leave events over a rolling time window. \textsf{``On-boarding trust"} is gathered directly from the system's sandbox environment, capturing the exact ratio of agents that pass automated behavioral vetting scripts prior to deployment. \textsf{``Policy consistency"} is measured by parsing identity and access management (IAM) audit logs to verify whether the timestamp of every access control list (ACL) update falls strictly within a predefined latency deadline. A low group volatility, high on-boarding trust, and perfect policy consistency together indicate that dynamic grouping does not degrade a system's trust boundary.}

\para{(C2's Metrics)} 
\revise{We quantify covert coordination risk through statistical anomaly detection and specialized classifiers. \textsf{``Collusion risk"} is calculated by continuously monitoring the agents' payoff surplus against a historic baseline distribution to flag mathematically improbable cooperation. The \textsf{``Covert-channel score"}~\cite{NEURIPS2024_861f7dad} generates concrete numbers by running sampled inter-agent message payloads through format compliance checkers and auxiliary steganography-detection models that output definitive confidence scores. The \textsf{``independence"} ratio simply parses execution logs to count the proportion of localized decisions made without calling peer agents. A rising collusion risk or covert-channel score paired with a falling independence ratio warns that a subset of agents may be manipulating outcomes.}

\para{(C3's Metrics)} \revise{We capture structural alignment quantitatively by comparing objective functions. \textsf{``Goal completeness"} is computed in the reinforcement learning setup by tracking the moving average of individual agent reward signals divided by the global environment reward. \textsf{``Conflict resolution"} is calculated by querying the arbitration module's logs to count the total number of flagged conflicting actions, and dividing it by the fraction that subsequently resulted in a documented, resolved state. \textsf{``Mutual benefit"} calculates the measured proportion of agents whose individual task payoffs strictly fulfill their baseline utility. Together, these metrics reveal whether reward shaping truly harmonizes individual and collective objectives.}

\para{(C4's Metrics)} \revise{We measure live model evolution via fine-tuning pipeline telemetry. \textsf{``Tuning log coverage"} counts the strict ratio of deployed parameter updates that possess a corresponding, cryptographically signed audit log. \textsf{``Drift-detection latency"} is obtained via continuous red-teaming: injecting synthetic anomalies into the network and measuring the exact timestamp delta (in seconds) until the anomaly detection module fires. \textsf{``Performance consistency"} is measured by automatically executing a standard static benchmark suite after every self-tuning cycle and recording the absolute pass rate.}

\para{(C5's Metrics)} \revise{We turn accountability into a measurable property by auditing the data pipeline. \textsf{``Provenance coverage"} and \textsf{``action traceability"} are computed by randomly sampling logged multi-agent outputs and running an automated tracing script; the metric is the ratio of queries that successfully reconstruct the full causal source-to-sink graph in the database. \textsf{``Source verification"} is measured at the API gateway as a strict count of cross-domain messages carrying a valid cryptographic signature divided by the total cross-domain message volume.}

\para{(C6's Metrics)} \revise{We instrument text-level defenses using evaluation datasets. A high \textsf{``Ill-prompt block rate"} and minimal \textsf{``false-positives"} are concretely measured by periodically routing a standardized dataset of known attacks and benign prompts through the semantic firewall, yielding exact classification ratios. \textsf{``Infection propagation"}~\cite{peignelefebvre2025multi} is measured in sandbox simulations by injecting an attack prompt and counting the precise number of downstream agents whose memory state or subsequent outputs reflect the injected payload.}

\para{(C7's Metrics)} \revise{We extend classical CIA principles to multi-LLM pipelines via network traffic analysis. \textsf{``Secure-channel utility"} is the ratio of cross-domain API calls using mutually authenticated transport-layer security (e.g., mTLS). \textsf{``Data leakage"} incidence and \textsf{``Request vetting"} are quantified by deploying a Data Loss Prevention (DLP) scanner over egress traffic logs to count the exact number of policy violations or illicit requests blocked per thousand queries.}

Because every metric above is a ratio, providers can stream them to a dashboard and set policy thresholds (e.g., ``halt execution if some metric drops below 0.9"). Researchers can report the same values to make security claims reproducible, and regulators gain a ready-made scorecard for certification audits---all from a single, compact table.

\subsection{Countermeasures}
\label{subsec:countermeasure}

.We provide possible directions to address each of the seven security issues. Some proposals (C3, C4) use a trusted arbitration system between domains, which is different from an hierarchical authority.

\para{(C1's Countermeasure) Trust-adaptive Dynamic Teaming: } 
Extending the idea of multi-agent partner selection~\cite{Anastassacos_Hailes_Musolesi_2020}, each agent can maintain a differentiable trust ledger -- a graph-neural module that updates peer-trust scores from message consistency, error rates, and rule violations observed in real time. Dialogue routes are weighted by these scores: low-trust peers are quarantined or demoted; high-trust peers gain routing priority or temporary leadership. Reinforcement learning drives the ledger toward team-level reward (task success + security bonus for blocking anomalies). Agents exchange signed ``trust snapshots,'' letting newcomers bootstrap without disclosing private logs.

\para{(C2's Countermeasure) Adversarial Multi-agent Training for Collusion Resistance: } A collusion generator can produce synthetic teaming scenarios where subsets of agents coordinate hidden agendas (via steganography, role laundering, etc.). Honest agents and a learnable incentive module co-train against these adversaries: the module maps conversation features into adaptive bonus/penalty signals, rewarding evidence-backed dissent and penalizing correlated misreports. Training proceeds in self-play cycles until Nash-like equilibrium where collusion yields no net gain. For example, ~\cite{brero2022learning} uses an adversarial training technique to prevent e-commerce pricing collusion among sellers. 

\review{[Reviewer 1] Regarding countermeasures, both C3 and C6 seem to call for a centralized component, which is a contradiction in terms, considering the scenario created for the purpose of flexible and distributed cooperation of autonomous AI agents. Its effectiveness is questionable, if the full details of the different domains the agents belong to are not disclosed. Even if appropriate, it would be a bottleneck.}

\para{(C3's Countermeasure) Hierarchical Conflict Arbitration via Meta-LLM Controller: } \revise{Domains can deploy decentralized local meta-controllers that communicate via a federated conflict-arbitration protocol. Each local meta-agent continually monitors its own lower-level agents' proposed actions. When cross-domain coordination is required, these paired controllers exchange limited, abstract representations of their goals rather than raw, proprietary directives. They use localized conflict-detection mechanisms to identify contradictions. Upon detecting a conflict, the paired meta-LLMs negotiate a resolving instruction. To maintain strict domain sovereignty, this instruction can be applied after being approved by human operators from both domains. Over time, subordinate agents can be fine-tuned locally using feedback from their respective meta-controllers, learning to preemptively avoid inconsistencies without relying on a central bottleneck.}


\para{(C4's Countermeasure) Cross-domain Reward Alignment via Adaptive Credit Assignment: } We can introduce an adaptive credit assignment network that aligns local rewards of each agent with the global multi-agent objective. During training, a shared critic model (potentially an LLM-based evaluator) observes the collective outcome and each agent’s contributions (e.g. action traces or textual outputs). It then computes tailored reward signals for each agent by estimating their marginal impact on the overall result. The system would dynamically adjust each agent’s reward function via gradient signals so that maximizing individual rewards also maximizes the team’s performance. .The adjusted reward plans can be vetted via a simulation or training before being deployed. While existing works~\cite{ferret2020selfattentional} focus on adaptively transferring a reward function from one task to another, our novel challenge is to train a differentiable communication channel among agents to propagate global reward information back into each agent’s policy update.

\para{(C5's Countermeasure) Neural Provenance Tracking with Embedded Signatures: } Each agent can embed subtle signatures (e.g., watermark~\cite{NEURIPS2024_65ec8f5c}) into its generated content to mark its contributions. Concretely, agents are trained to include quasi-imperceptible metadata in their textual outputs – for example, by biasing certain token choices or using a private vocabulary of markers – that do not alter the apparent meaning but carry identifying information. A dedicated decoder model or forensic LLM can later extract these hidden signatures from the final multi-agent output, reconstructing a timeline or graph of which agent produced or modified each part. One challenge is to harden this neural signature scheme to survive transformations (e.g. rephrasing by subsequent agents) and to remain robust against adversarial removal.

\para{(C6's Countermeasure) Session-level Semantic Firewalls: } 
Instead of turn-by-turn filtering, a dedicated firewall LLM ingests the entire multi-agent dialogue in a sliding window, building a contextual knowledge graph of entities, quantitative tokens, and policy tags (e.g., ``salary-individual''). \revise{Note that each domain manages its own firewall in a decentralized manner. Each domain's firewall enforces its own policies with full local visibility and determines whether each session violates the session-level semantic rules; residual composite leaks arise only when individual responses are each policy-compliant yet jointly revealing, which is mitigated by conservative policy design that accounts for external combinability.} It uses contrastive training to spot when a new utterance, combined with prior context, breaches a policy. Detected violations trigger automatic redaction or request-for-clarification messages. The firewall’s policy model continually fine-tunes on anonymized leak/non-leak transcripts shared across domains under differential privacy.


\para{(C7's Countermeasure) Verifiable Reasoning with Privacy: } In blind inference, an agent emits (i) an encrypted answer and (ii) a public proof sketch---hash-bound logic traces or statistical bounds---letting a verifier LLM confirm semantic correctness without seeing the private input. Lightweight ZKPs~\cite{NEURIPS2024_65ec8f5c} and encrypted-embedding checks (run in secure enclaves) enable this, yet current protocols remain too slow for real-time use. A potential fix is a hybrid design that blends these proofs with fast two-party linear MPC~\cite{10.1007/978-3-319-70694-8_22}, slashing latency while preserving privacy.

%% file: 050-conclusion.tex
\section{Conclusions and Outlook}
\label{sec:conclusion}

Cross-domain multi-agent LLMs hold transformative promise, but only if security becomes a first-class design constraint rather than an afterthought. By mapping seven unaddressed challenge areas, we have shown that neither single-agent defenses nor traditional multi-agent safeguards suffice once models cross ownership boundaries. Tackling these problems demands tight collaboration between the AI-safety, cryptography, and distributed-systems communities, coupled with rigorous open benchmarks to quantify the security–utility trade-offs unique to cross-domain deployments. 
The research directions we outline argue for security primitives that are as adaptive and learning-centric as the agents themselves. 
Addressing them now will prevent tomorrow's agent societies from repeating the Internet's costly security debt.



